\documentclass[final]{elsart}

\usepackage{amssymb}
\usepackage{amsmath}
\usepackage{bm}
\usepackage{graphicx}
\usepackage{subfigure}
\usepackage{natbib}
\journal{Atmospheric Environment}

\begin{document}

\begin{frontmatter}

\title{Environmental  Time Series Interpolation Based on Spartan Random Processes }

\author[Tuc1]{Milan \v{Z}ukovi\v{c}}
\ead{mzukovic@mred.tuc.gr}
\author[Tuc1]{D. T. Hristopulos \corauthref{cor}}
\ead{dionisi@mred.tuc.gr} \corauth[cor]{Corresponding author.
Telephone: +30-28210-37688. Fax: +30-28210-37853.}
\address[Tuc1]{Technical University of Crete, Geostatistics Research
Group, Department of Mineral Resources Engineering, Chania, 73100,
Greece}
%

\begin{abstract}

In many environmental applications, time series are either
incomplete or irregularly spaced. We investigate the application of
the Spartan random process to missing data prediction. We employ a
novel modified method of moments (MMoM) for parameter inference. The
CPU time of MMoM is shown to be much faster than that of maximum
likelihood estimation and almost independent of the data size. We
formulate an explicit Spartan interpolator for estimating missing
data. The model validation is performed on both synthetic data and
real time series of atmospheric aerosol concentrations. The
prediction performance is shown to be comparable with that attained
by the best linear unbiased (Kolmogorov-Wiener) predictor at reduced
computational cost.
\end{abstract}



\end{frontmatter}

\section{Introduction}

Time series have wide-ranging applications in environmental
monitoring, such as air and water quality control.
The series carry information about temporal autocorrelations
(henceforth, correlations) in variables such as atmospheric
pollutant concentrations, particulate matter, indicators of water
clarity, salinity etc.
%
Knowledge of  the correlation structure enables the prediction of
time-series, estimating the prediction uncertainty, and developing
stochastic simulations that reconstruct (at least partially) the
process of interest. In many applications, observations are
irregularly spaced (due to difficulties and cost of data
acquisition), and they include fully missing (due to measurement
interruptions) or censored (lying below the equipment detection
threshold) data.  These gaps in the series prevent the use
of standard analytical techniques that assume regular sampling.
Recently, the European Union introduced a novel concept in
environmental legislation, in an effort to establish common data
quality objectives for air pollution \citep{ball05}. Hence, there is
a need to harmonize data obtained by means of different measurement
techniques and sampling conditions. Difficulties in quantifying
uncertainty, due to incomplete time coverage, led to simplistic
approaches that assume a random distribution of the missing data.
The associated uncertainty is then reduced to a simple function of
the percentage of the missing data. However, neglecting the
frequency distribution and temporal correlations of the missing data
can strongly  affect the estimates. Furthermore, the data may be
sampled at different temporal scales. By predicting time series at
unmeasured instants, it is possible to fill in the missing values or
to down-scale (refine) existing measurements.

Several recent papers evaluate methods of time series prediction for
environmental applications. In \citep{house05}, a first-order
autoregressive model was applied to irregularly spaced water clarity
data. Multivariate models have been applied to the time series of
pollutant concentrations in the Arctic, using the multiple
imputation approach \citep{hopke01}. Time series of aerosol particle
number concentrations were modeled on the basis of traffic-related
air pollution and meteorological variables \citep{paat05}. In
\citep{junni04}, the authors evaluate and compare various univariate
and multivariate methods for missing data imputation in air quality
data sets. Generally, the simple univariate models that either
utilize only local information or make linearity assumptions, are
fast but their scope is limited. Multivariate models provide higher
accuracy and reliability at increased computational cost. In the
current study, we present a novel linear predictor, which is based
on the use of ``pseudo-energy'' functionals inspired from
statistical physics. We apply this Spartan predictor in conjunction
with  computationally efficient parameter inference method.  We show
that the  results obtained with the Spartan predictor are comparable
with those obtained by the best linear unbiased estimator. However,
the former is superior in terms of computational speed.

For Gaussian time series, the temporal structure is determined from
the autocovariance (henceforth, covariance) matrix, which is
estimated from the data. In the case of non-uniform sampling steps,
the structure function (variogram) is typically estimated instead of
the covariance for practical reasons. However, temporal correlations
are also present (albeit on different physical scales) in models of
statistical physics, e.g., in the Ising model and spin glass models.
In these models, correlations are imposed by means of physical
interactions embodied in the energy functional and thus do not need
to be calculated from the data.
Recently, the method of Spartan Spatial Random Fields (SSRF)
\citep{dth03,dthel07} was proposed as a general framework for
geostatistical applications. SSRFs are parametrically flexible, do
not rely on variogram estimation to determine the spatial structure,
and allow incorporating physical constraints in the joint
probability density function. In the present study, we define in the
same spirit the Spartan Random Processes (SRP) and apply them to
 time series prediction.

The rest of the paper is organized as follows.  Spartan random
processes are introduced  in Section \ref{spartan}. In Section
\ref{parinf}, we discuss parameter inference using  maximum
likelihood estimation and the modified method of moments. In Section
\ref{spatest} we present the  Spartan interpolator and compare it to
the Kolmogorov Wiener predictor (simple kriging). In
Sections~\ref{simul} and~\ref{sec:real},
 we compare the model inference and
 data prediction methods using both synthetic data
 as well as real time series of aerosol concentration.
 Finally, we summarize and present our conclusions in Section \ref{concl}.

\section{Spartan Random Processes}
\label{spartan} Herein we assume a Gaussian, second-order stationary
\citep{yaglom}, detrended time series, $X_{\lambda}(t);$ $\lambda$
is an intrinsic time scale related to temporal resolution.
%
The pdf can be expressed in terms of an energy functional $
{H[X_\lambda({t})]} $, according to the familiar from statistical
physics expression of Gibbs pdf's:
\begin{equation}
\label{gibbspdf} f_{\rm x} [X_\lambda] = Z^{- 1} \, e^{-
H[X_\lambda(t)]},
\end{equation}
where the partition function $ Z $ is the normalization factor. In
the classical geostatistical framework the energy functional
corresponds to:
\begin{equation}
\label{covenergy} H[X_\lambda] = \frac{1}{2} \sum_{i=1}^{N}
\sum_{j=1}^{N}  X_\lambda (t_i)\, [ G_{\rm x}]^{ - 1}_{i,j}
X_\lambda (t_j),
\end{equation}
where $ [ G_{\rm x}]^{ - 1}_{i,j} $  is the inverse of the
covariance matrix (also known as the precision matrix), and $N$ is
the number of the data points.

In analogy with Spartan spatial random fields~\citep{dth03}, we
define the fluctuation-gradient-curvature (FGC) Spartan random
process that involves four parameters with a well defined physical
meaning: the scale coefficient $\eta_0$, the shape coefficient
$\eta_1$, the characteristic time $\xi$, and the cutoff circular
frequency $k_c \propto \lambda ^{-1}$. A kernel function is used to
implement the cutoff. Below we use a boxcar kernel with sharp
spectral cutoff at $k_c.$
If time is considered as a continuous variable, the FGC Spartan pdf
is determined from the following energy functional:
\begin{equation}
\label{fgc_1D} H_{\rm fgc} [X_\lambda ; \bm{\theta} ] =
\frac{1}{{2\eta _0 \xi }}\int_{- \infty}^{\infty} {dt} \, \left\{
\left[ {X_\lambda (t)} \right]^2 + \eta _1 \,\xi ^2 [
{\dot{X}_\lambda (t)} ]^2  + \xi ^4 [ {\ddot{X}_\lambda (t)}]^2
\right\},
\end{equation}

\noindent where $\bm{\theta}=(\eta_0,\eta_1,\xi,k_c)$ and  the dots
denote the first and second order time derivatives. The covariance
spectral density is then given by the following expression
\begin{equation}
\label{covspd1} \tilde{G}_{\rm x}(k; \bm{\theta})=
 \frac{ h_\lambda (k)  \,\eta _0 \,\xi  }{1 + \eta _1 \,(k \xi )^2  + (k \xi )^4 },
\end{equation}

\noindent where $h_\lambda  (k) =1,$ if $k \le k_c$ and  $h_\lambda
(k) =0$ otherwise. The covariance function is then obtained from the
inverse Fourier transform, given by the following integral:

\begin{equation}
\label{cov1d} G_{\rm x}(\tau; \bm{\theta})={\int}_{-\infty}^{\infty}
\frac{dk}{2\pi} \,
              \tilde{G}_{\rm x}(k; \bm{\theta})\, e^{\jmath k \tau}.
\end{equation}

For a discrete time series, sampled at the times $t_{n}=n \alpha$,
$n=1,\ldots,N$, $\alpha >0$, the derivatives are approximated by
finite forward differences and the energy functional takes the
following form:

\begin{equation}
\label{H1d} H_{\rm fgc} \left[ X_{\lambda};{\bm \theta} \right]   =
\frac{1}{2\eta _0 \xi }
  \sum\limits_{n = 1}^{N} \left\{  S_{0}( t_{n})
  + \eta _1 \, \xi ^{2} S_{1}(t_{n})
  + \xi^{4}  S_{2}(t_{n}) \right\},
\end{equation}

\noindent where $$S_{0}(t_n)= X_{\lambda} (t_n )^{2},$$
$$S_{1}(t_n)= \left[ \frac{ X_{\lambda} (t_n + \alpha) - X_{\lambda}
(t_n) }{\alpha} \right]^{2},$$  $$S_{2}(t_n)=  \frac{
\left[X_{\lambda} (t_n + \alpha) + X_{\lambda} (t_n-\alpha) -2
X_{\lambda}(t_n)\right]^{2} }{\alpha^{4}} .$$


\subsection{The Precision Matrix}
\label{ssec_prec-mat} In Eq.~(\ref{covenergy}), the values of the
process are coupled even at distant times through the precision
matrix. In contrast, in Eq.~(\ref{H1d}) only values between
neighboring times are coupled. The  energy functional is then
expressed as follows:
\begin{equation}
\label{hatH} H_{{\rm{fgc}}}\left[ X_{\lambda} ; \bm{\theta}
\right]=\frac{1}{2} X_{\lambda}(t_i)J_{\rm x}(t_i,t_j;
\bm{\theta})X_{\lambda}(t_j),
\end{equation}
where $J_{\rm x}(t_i,t_j; \bm{\theta})$ is the {\it{precision
matrix}}. Based on Eq. (\ref{H1d}), the precision matrix can be
expressed in closed form as follows:
\begin{equation}
\label{pJ} J_{{\rm{x}}}(t_i,t_j; \bm{\theta})=\frac{1}{\eta_0 \xi}
\Bigg\{J_0(t_i,t_j)+\eta_1
\frac{\xi^2}{\alpha^2}J_1(t_i,t_j)+\frac{\xi^4}{\alpha^4}J_2(t_i,t_j)\Bigg\},
\end{equation}
\noindent where $J_0(t_i,t_j)=\delta_{i,j}$ is the identity matrix,
$ J_1(t_i,t_j)$ is the gradient precision sub-matrix,
\begin{equation}
\label{eq:J1} J_1(t_i,t_j)= \left(\begin{array}{ccccc}
1 & -1 & 0 & \cdots & 0\\
-1 & 2 & -1 & \cdots & 0\\
   & \ddots & \ddots & \ddots &  \\
0 & \cdots & -1 & 2 & -1\\
0 & \cdots & 0 & -1 & 1\\
      \end{array} \right),
\end{equation}
and $ J_2(t_i,t_j)$ is the curvature precision sub-matrix
\begin{equation}
\label{eq:J2} J_2(t_i,t_j)= \left(\begin{array}{cccccc}
1  &  -2  &  1  &  0  & \cdots & 0\\
-2 &   5  & -4  &  1  & \cdots & 0\\
1  &  -4  &  6  & -4  & 1      & 0\\
   & \ddots & \ddots & \ddots & \ddots &  \\
0  & \cdots & 1 & -4 &  5 & -2\\
0  & \cdots & 0 &  1 & -2 &  1\\
     \end{array} \right).
\end{equation}
In the discrete representation, given by
Eqs.~(\ref{hatH})-(\ref{eq:J2}), the Spartan random process is
equivalent to a Markov random process, the only difference from the
latter being the nonlinear dependence on the coefficients. However,
in contrast with Markov processes, the Spartan random process can be
generalized to non-uniform sampling patterns as shown
in~\citep{serra08}.

\section{Parameter Inference}
\label{parinf} Let $T_{\rm s}=\{t_1,...,t_N\}$ be a set of sampling
times and ${\bf X}^{*}(T_{\rm s}) =\{X^{*}_{1},...,X^{*}_{N}\}$ the
vector of sample measurements. Let $L(\bm{\theta}|{\bf X}^{*})$
denote the likelihood of the parameter vector $\bm{\theta}$ given
the data. Let us also define the reduced parameter set
$\bm{\theta'}=(\eta_1,\xi,k_c)$ and the scaled precision matrix $
J'_{\rm x}(\bm{\theta'}) =  \eta_0 \, J_{\rm x}(\bm{\theta}).$ Let
the sample estimate of the energy functional be $ \hat{H}_{\rm{fgc}}
\left[ X_{\lambda}; \bm{\theta} \right]$. The scaled energy
functional is defined by $\hat{H'}_{\rm{fgc}}\left[ X_{\lambda};
\bm{\theta'} \right] = \eta_0 \, \hat{H}_{\rm{fgc}} \left[
X_{\lambda}; \bm{\theta} \right]$.

For a complete (with no missing data) series $ \hat{H}_{\rm{fgc}}
\left[ X_{\lambda}; \bm{\theta} \right]$  is given from
equation~(\ref{hatH}) by replacing $X_{\lambda}(t_i)$ with the data
$X^{*}(t_i).$

In the case of series with missing data it is preferable to use the
definition
\begin{equation}
\label{eq:estim-H1d} \hat{H}_{\rm fgc} \left[ X_{\lambda};{\bm
\theta} \right] = \frac{N}{2\eta _0 \xi }
   \left\{  \overline{S_{0}( t)}
  + \eta _1 \, \xi ^{2} \overline{S_{1}(t)}
  + \xi^{4}  \overline{S_{2}(t)} \right\},
\end{equation}
where $\overline{S_{q}}$ are sample estimates of the respective
quantities $S_{q}, \, q=0,1,2$, evaluated over all the time instants
present in the series. The gaps do not influence the estimation  of
$\overline{S_0}$, which is performed over all the sample points. In
the case of $\overline{S_{1}},$ the average involves all compact
clusters of data that include at least two nearest neighbors.
Similarly, in the case of $\overline{S_{2}},$ the average involves
all compact clusters of data that include at least three nearest
neighbors.

\subsection{Maximum Likelihood Estimation}
\label{mle} The maximum likelihood estimates (MLE) are obtained by
minimizing numerically the negative log-likelihood (NLL),
e.g.~\citep*[pp. 169-175]{stein99}, which requires the evaluation
and inversion of the covariance matrix. The numerical operations are
computationally intensive, especially for large sample sizes. As we
show below, the Spartan random process has a significant advantage
in terms of computational speed. The NLL becomes:

\begin{equation}
\label{nnl1} -\log L(\bm{\theta}|{\bf X}^{*})=
\frac{\hat{H'}_{\rm{fgc}} \left[ X_{\lambda}; \bm{\theta'}
\right]}{\eta_0} + \frac{N}{2}\log (\eta_0) + \frac{N}{2}\log (2\pi)
- \frac{1}{2}\log |J'_{\rm x} (\bm{\theta'})|.
\end{equation}
The estimate $\hat{\eta}_0$ follows from requiring the derivative of
$\log L(\bm{\theta}|{\bf X}^{*})$ with respect to $\eta_0$ to
vanish, leading to,
\begin{equation}
\label{eta0} \hat{\eta}_0=\frac{2\hat{H'}_{\rm{fgc}}\left[
X_{\lambda}; \bm{\theta'} \right]}{N},
\end{equation}
and by replacing $\eta_0$ with $\hat{\eta}_0$ in
equation~(\ref{nnl1}), the NLL is given in terms of the scaled
variables $\hat{H'}_{\rm{fgc}}\left[ X_{\lambda}; \bm{\theta'}
\right]$ and $ J'_{\rm x}(\bm{\theta'})$ as follows:
\begin{equation}
\label{nnl2} -\log L(\bm{\theta}|{\bf X}^{*})=\frac{N}{2}
\log\bigg({2\hat{H'}_{\rm{fgc}}\left[ X_{\lambda}; \bm{\theta'}
\right]/N}\bigg)-\frac{1}{2}\log |J'_{\rm x}(\bm{\theta'})|+ C_N.
\end{equation}

\noindent The constant $C_N=\frac{N}{2}[\log (2\pi)+1]$ is
independent of $\bm{\theta'}$ and can be dropped from the
minimization.  The NLL is minimized using a numerical optimization
method. In the case of the Spartan random process, the computational
efficiency is gained by the fact that $\hat{H'}_{\rm{fgc}}\left[
X_{\lambda}; \bm{\theta'} \right]$ is estimated  without the
inversion of a full covariance matrix. Moreover, based on
equations~(\ref{eq:J1}) and~(\ref{eq:J2}), $J'_{\rm
x}(\bm{\theta'})$ is a pentadiagonal symmetric matrix, so that  fast
and accurate approximations can be used for the evaluation of its
determinant~\citep{reusk02}.

\subsection{Modified Method of Moments}
\label{sec:monc}

The modified method of moments (MMoM) is based on fitting stochastic
constraints with their  sample counterparts. The constraints are
based on short-range correlations and are motivated from the terms
$S_{m}(t_{n}),\ m=0,1,2$, in the energy functional~(\ref{H1d}).
 The stochastic constraints are expressed as follows:
\begin{equation}
\label{s01d} E[S_0]=G_{\rm x}(0),
\end{equation}
\begin{equation}
\label{s11d} E[S_1]=\frac{2}{\alpha^2} \left[ G_{\rm x}(0) -G_{\rm
x}(\alpha) \right],
\end{equation}
\begin{equation}
\label{s21d} E[S_2]=\frac{2}{\alpha^4} \left[ 3G_{\rm x}(0) + G_{\rm
x}(2\alpha) -4 G_{\rm x}(\alpha)\right].
\end{equation}
The stochastic constrains are functions of ${\bm \theta}$ obtained
from spectral integrals~\citep{dth03}.
%
%
%
Herein we focus on time series with well defined temporal resolution
and correlation times that exceed this resolution. Then, we can
assume an infinite band limit ($k_c \rightarrow \infty$) and
suppress the subscript $\lambda$.  The respective Spartan covariance
function has been evaluated in~\cite{dthel07}, and it is given by
the following:

\begin{equation}
\label{eq:spart-cov} G_{\rm x}(t)= \left\{
\begin{array}{ll}
\eta_{0}e^{-h\beta_2}\bigg[\frac{\cos(h\beta_1)}{4\beta_2} +
\frac{\sin(h\beta_1)}{4\beta_1}\bigg],         & {\rm{for}}\ |\eta_1| < 2, \\
\eta_{0}\frac{(1+h)}{4e^{h}},                             & {\rm{for}}\ \eta_1 = 2, \\
\eta_{0}\frac{1}{\Delta}\bigg[\frac{e^{-h\omega_1}}{2\omega_1} -
\frac{e^{-h\omega_2}}{2\omega_2}\bigg], & {\rm{for}}\ \eta_1 > 2,
\end{array} \right.
\end{equation}
where $h \equiv |\tau|/\xi$, $\beta_{1,2} =( |2 \pm
\eta_1|)^{1/2}/2$, $\omega_{1,2} = (|\eta_1 \pm \Delta|/2)^{1/2}$,
and $\Delta = |\eta_1^2 - 4|^{1/2}$. Hence, the stochastic
constraints given by Eqs. (\ref{s01d}-\ref{s21d}) can be expressed
analytically in terms of the covariance function. The optimal values
of the model parameters are estimated by minimizing the following
objective functional \citep{dth03}

\begin{equation}
\label{dm} \Phi_{s}[X(t)]=\Bigg |1-\sqrt{\frac{\overline{S_{1}}}
{\overline{S_{0}}}\frac{E[S_{0}]}{E[S_{1}]}}\Bigg|^2+
\Bigg|1-\sqrt{\frac{\overline{S_{2}}}{\overline{S_{1}}}\frac{E[S_{1}]}{E[S_{2}]}}\Bigg|^2
\end{equation}
The application of this distance metric (DM) is based on the
following assumptions: First, the sample averages
$\overline{S_{m}}$, $m=0,1,2$ are accurate and precise estimators of
the stochastic expectations $E'[S_{m}]$ of the underlying random
process (where $E'[.]$ denotes the expectation with respect to the
unknown probability density). This assumes that ergodic conditions
are satisfied, e.g.~\citep{adler}. Second, the stochastic
expectations $E[S_{m}]$ of the Gibbs random process should
approximate the expectations $E'[S_{m}]$ of the underlying process.

\section{Prediction of Missing Data (Temporal Interpolation)}
\label{spatest} Let us assume that there are points where data are
missing in the sampling set $T_{\rm s}.$ The missing points are
included in the prediction set, $T_{\rm p}=\{z_1,...,z_P\}$, which
is assumed to be disjoint from $T_{\rm s}$.  We denote by  $\hat{\bf
X}(T_{\rm p})$ the vector of estimates (temporal predictions).
Finally, let $T=T_{\rm s} \cup T_{\rm p}$ and ${\bf X}(T)= \hat{\bf
X}(T_{\rm p}) \circ {\bf X}^{*}(T_{\rm s})$ be the joint vector of
measurements and prediction points.

Using the  Kolmogorov-Wiener theory~\citep{kitan,wack}, the
single-point prediction at the point $z_p,$ $\hat{X}(z_{p}),$ is
given  as a linear superposition of the data values. The
coefficients of the superposition are selected so as to ensure zero
bias and minimize the mean square error of the prediction, leading
to the following Kolmogorov-Wiener prediction (KWP) equation:
\begin{equation}
\label{krig} \hat{X}(z_p) = \left[ {\bf G}^{-1}({\bm \theta}; T_{\rm
s},T_{\rm s}) \, {\bf G}({\bm \theta}; T_{\rm s},z_{p}) \right]^{\rm
tr} \, {\bf X}^{*}(T_{\rm s}), \ \ \ p=1,...,P.
\end{equation}
In the above, ${\bf G}({\bm \theta}; T_{\rm s},T_{\rm s})$ is the $N
\times N$ data covariance matrix, ${\bf G}^{-1}$ is its inverse,
${\bf G}({\bm \theta}; T_{\rm s},z_{p})$ is the $N \times 1$
covariance matrix between the estimated point and the data, and
${\bf A}^{\rm tr}$ denotes the transpose of the matrix ${\bf A}.$

The Spartan family of covariance functions,
Eq.~(\ref{eq:spart-cov}),  can be used in KWP algorithms to provide
new types of spatial dependence. Within the SRP framework it is also
possible to define a new type of linear predictor, which allows
multiple-point prediction to be performed simultaneously over all
points in $T_{\rm p}$.

\subsection{The Spartan predictor}
\label{sing-point}

The Spartan predictor (SP) defined below, is based on the FGC model.
 It relies on maximizing the conditional  probability density, $f_{\rm x} [\hat{\bf X}(T_{\rm p})
|{\bf X}^{*}(T_{\rm s}) ]$. Considering the definition $f_{\rm x}
[\hat{\bf X}(T_{\rm p}) | {\bf X}^{*}(T_{\rm s}) ] = f_{\rm x}
[\hat{\bf X}(T_{\rm p})]/f_{\rm x} [{\bf X}^{*}(T_{\rm s}) ]$, the
problem reduces to maximizing $f_{\rm x} [{\bf X}(T)]$. The latter
involves the energy functional over the combined set of sampling and
prediction points, i.e.,
\begin{equation}
\label{hatH1} \hat{H}_{\rm fgc}\left[ {\bf X}(T) ; \bm{\theta}
\right]= \frac{1}{2} {\bf X}^{\rm tr}(T) \, {\bf J}_{\rm x}
(\bm{\theta}) \, {\bf X}(T).
\end{equation}

\noindent Maximizing the conditional probability with respect to the
prediction values leads to the following linear system of $P$
equations

\begin{equation}
\label{lin_eqs} \frac{\partial \hat{H}_{fgc}[ {\bf X}(T);
\bm{\theta} ]} {\partial X(z_p)}\Bigg |_{\hat{X}(z_p)} = 0, \ \ \
p=1,\ldots,P.
\end{equation}
The predictions do not depend on $\eta_0$, since the latter is an
overall scaling factor for ${\bf J}_{\rm x} (\bm{\theta})$. If there
are no {\it{interactions}} between the prediction points, i.e.
$J_{\rm x} ({\bm \theta}; z_i,z_j)=0$, $i,j=1,...,P$, the linear
predictor is expressed explicitly by

\begin{equation}
\label{lin_pred} \hat{X}(z_p) = -\sum_{ t_{l} \in V(z_p)}
{\frac{J_{\rm x}({\bm \theta}; t_{l},z_p)} {J_{\rm x}({\bm \theta};
z_{p},z_{p})} \, X^{*}(t_{l})}, \ \ \ p=1,...,P,
\end{equation}
where $V(z_p)$ is the interaction neighborhood of $z_p$, i.e., the
set of  the points in $t_{l} \in T_{\rm s}$ that interact with
$z_p$. In the case of the FGC functional the interaction
neighborhood spreads up to the second-nearest neighbor.

\begin{figure}[htbp]
  \begin{center}
\rotatebox{0}{ \resizebox{10cm}{!}{
\includegraphics{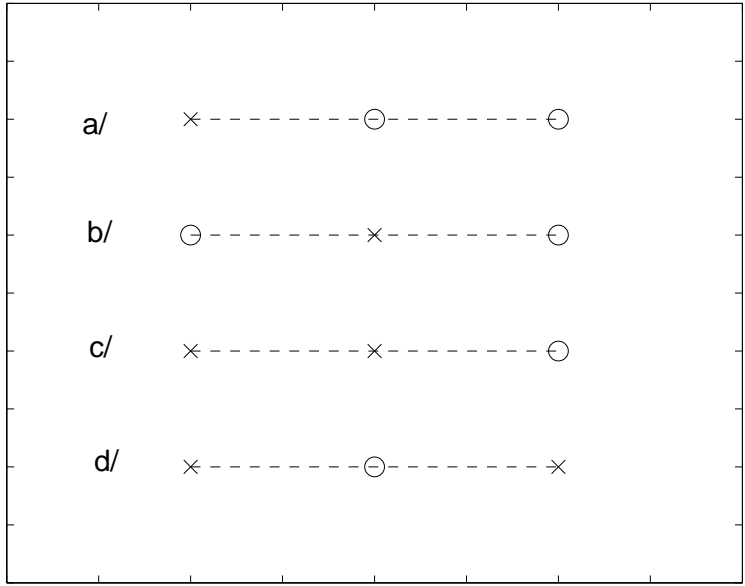}}}
  \end{center}
\caption{Some trivial time series, consisting of only 3 points,  for
which equivalence between SP and KWP can be shown analytically. The
circles denote the known data and the crosses the prediction
points.} \label{fig:elem_chains}
\end{figure}

The Spartan predictor is linear and unbiased. Unlike the KWP, it
 does not require computationally expensive calculations of the
covariance matrix. Specifying an arbitrary search neighborhood for
each prediction point is not necessary, since the SP uses only the
data in the  immediate interaction neighborhood. In KWP the
estimation is performed sequentially point-by-point by solving a
linear system of equations. In contrast, SP is a multi-point
estimator and it can be formulated explicitly only if the
interaction neighborhoods of the estimation points are disjoint. The
numerical complexity of multipoint SP estimation is $O(P^3)$, while
the numerical complexity of KWP is $O(P\, M^3)$, where $M$ is
average number of points inside
 the local search neighborhood of the estimation points.
In some trivial cases (Fig. \ref{fig:elem_chains}), the equivalence
between the FGC SP and KWP can be shown analytically, using symmetry
of the precision and covariance matrices of the regularly spaced
data. For longer time series, the analytical comparison becomes
cumbersome and hence, we resort to comparison by numerical
calculations. In a general case, however, we cannot expect
equivalence of the two methods, since SP only considers the data
from the short-range interaction neighborhood while KWP considers
all the data available (or those from the search neighborhood,
generally different from the SP interaction neighborhood).
Nevertheless, in the multipoint Spartan prediction, the local
information propagates through the system via the interacting
prediction points, .i.e, system of coupled equations
(\ref{lin_eqs}), influencing predictions at distant locations and,
hence, its performance compared to KWP is a priori not obvious.

\section{Case Study I: Simulated Data}
\label{simul} All the computations were performed in the
Matlab$\circledR$ environment, on a desktop computer with Pentium 4
CPU at 3 GHz and 1 GB of RAM. We evaluate the performance of the MLE
and the MMoM methods for parameter inference. We also compare the
performance of the SP and KWP missing data estimators. First, we use
control time series of sizes $N=100,200,300,500,1000$ based on the
following covariance functions:  (i) Gaussian $G_{\rm x;G}
({\tau})=\sigma^{2} {\rm{exp}}(-h^{2}),$ (ii) exponential $G_{\rm
x;E} ({\tau})=\sigma^{2} {\rm{exp}}(-h),$ (iii) spherical $G_{\rm
x;S}({\tau})=\sigma^{2} (1-1.5h +0.5h^{3})$ for $h \le 1$ and
$G_{\rm x;S} ({\tau})=0$ for $h>1,$ and (iv) Whittle-Mat\'{e}rn
$G_{\rm x;M} ({\tau})=\sigma^{2}
\frac{{2}^{1-\nu}}{\Gamma(\nu)}(\kappa \tau)^{\nu}K_{\nu}(\kappa
\tau).$ In the above,  $\sigma^2$ represents the variance, $b$ the
correlation time, $h=|\tau|/b,$ the normalized time lag, $\nu$
 the smoothness parameter, $\kappa$ the inverse length, and $K_{\nu}$
the modified Bessel function of index $\nu$. The samples are
generated using the multivariate normal simulation method
\citep{johns87}.
The covariance parameters are: For the Gaussian $(\sigma,b)=(10,3)$,
for the exponential and the spherical $(\sigma,b)=(10,5)$, and for
the Whittle-Mat\'{e}rn: $(\sigma,\kappa,\nu)=(10,1,3.5)$. For each
covariance model and size $N$, 100 time series ${\rm X}^{*}_{i}$
($i=1,...,100$) are generated in order to calculate optimization and
computer time statistics.

\subsection{Results: Parameter Inference}
The optimization employs the Nelder-Mead simplex search algorithm
\citep{press} for both the MLE and the MMoM cases. This algorithm is
fast, because it does not require the computation of a Jacobian
matrix. The optimization is terminated when both the model
parameters and the objective function (NLL for the MLE or DM for the
MMoM) change between consecutive steps less than the specified
tolerance, $\epsilon = 10^{-6}$. The initial guesses for the Spartan
parameters are $\xi^{(0)}_{i}=\alpha=1$,
 $\eta_1^{(0)}=-1$ (for the MLE different initial guesses $\eta_1^{(0)}$ are
 in order due to multimodality).

In Table \ref{tab:par_inf_gauss},  we compare the MLE and MMoM
parameter inference methods, using time series with Gaussian
correlations. The mean values of the estimated Spartan parameters,
$\langle \eta_0^{*}\rangle$, $\langle \eta_1^{*}\rangle$ and
$\langle \xi^{*}\rangle$ are listed, as well as their standard
deviations $\hat{S}_{\eta_0^{*}}$, $\hat{S}_{\eta_1^{*}}$ and
$\hat{S}_{\xi^{*}}$. The number of the optimization iterations
$\langle N_{it} \rangle$, the optimization CPU time $\langle T_{cpu}
\rangle$, and the cost function mean value $\langle F^{*} \rangle$
($F^{*}=$ NLL for MLE and $F^{*}=$ DM for MMoM) at termination are
also tabulated.  The large standard deviations of $\eta_0$ and
$\eta_1$ for small $N$ (larger for the MMoM), decrease with
increasing $N$. The dependence on $\eta_1$ can be attributed to the
shape of the objective functions.

\begin{table}[h]
\caption{Spartan parameter inference by MLE and MMoM  (in the table
abbreviated as ML and MM, respectively) on 100 samples for each
domain size $N$ with the Gaussian covariance dependence (similar
results are obtained with the other models). The calculated
statistics: the mean values of the estimated parameters $\langle
\eta_0^{*}\rangle$, $\langle \eta_1^{*}\rangle$ and $\langle
\xi^{*}\rangle$, their standard deviations $\hat{S}_{\eta_0^{*}}$,
$\hat{S}_{\eta_1^{*}}$ and $\hat{S}_{\xi^{*}}$, the number of the
optimization iterations $\langle N_{it} \rangle$, the optimization
CPU time $\langle T_{cpu} \rangle$, and the cost function mean value
$\langle F^{*} \rangle$ at termination.} \label{tab:par_inf_gauss}
\begin{center}
\begin{tabular}{lrrrrrrrrrr}
\hline
           & \multicolumn{ 2}{c}{N=100} & \multicolumn{ 2}{c}{N=200} & \multicolumn{ 2}{c}{N=300} & \multicolumn{ 2}{c}{N=500} & \multicolumn{ 2}{c}{N=1\ 000} \\
\hline
    Par. &        ML &       MM &        ML &       MM &        ML &       MM &        ML &       MM &        ML &       MM \\
\hline
$\langle \eta_0^{*}\rangle$ &     80.95 &    125.70 &     88.08 &    130.82 &     88.35 &    135.66 &     87.26 &    133.67 &     89.52 &    134.91 \\
$\langle \eta_1^{*}\rangle$ &     $-$1.85 &     $-$1.52 &     $-$1.81 &     $-$1.54 &     $-$1.80 &     $-$1.53 &     $-$1.80 &     $-$1.54 &     $-$1.79 &     $-$1.53 \\
$\langle \xi^{*}\rangle$ &      2.01 &      1.91 &      2.10 &      1.98 &      2.12 &      1.99 &      2.13 &      2.01 &      2.15 &      2.03 \\
$\langle N_{it} \rangle$ &     42.98 &    151.19 &     42.30 &    149.85 &     41.70 &    149.34 &     41.94 &    152.47 &     41.33 &    153.36 \\
$\langle T_{cpu} \rangle$ &      0.46 &      0.08 &      1.59 &      0.08 &      4.01 &      0.08 &     12.94 &      0.08 &     82.28 &      0.08 \\
$\langle F^{*} \rangle$ &    104.3 &   $2$E$-21$ &    203.1 &   $1$E$-21$ &    301.1 &   $1$E$-21$ &    495.2 &   $7$E$-22$ &    990.9 &   $1$E$-21$ \\
$\hat{S}_{\eta_0^{*}}$ &     31.69 &     57.93 &     23.98 &     40.29 &     19.50 &     35.02 &     15.39 &     23.07 &     10.71 &     16.39 \\
$\hat{S}_{\eta_1^{*}}$ &      0.08 &      0.34 &      0.07 &      0.20 &      0.07 &      0.16 &      0.05 &      0.10 &      0.04 &      0.08 \\
$\hat{S}_{\xi^{*}}$ &      0.23 &      0.24 &      0.16 &      0.17 &      0.15 &      0.13 &      0.11 &      0.12 &      0.08 &      0.08 \\
\hline
\end{tabular}

\end{center}
\end{table}

\begin{figure}[htp]
  \begin{center}
    \subfigure[MMoM]{\label{fig:gauss_dm}\includegraphics[scale=0.3]{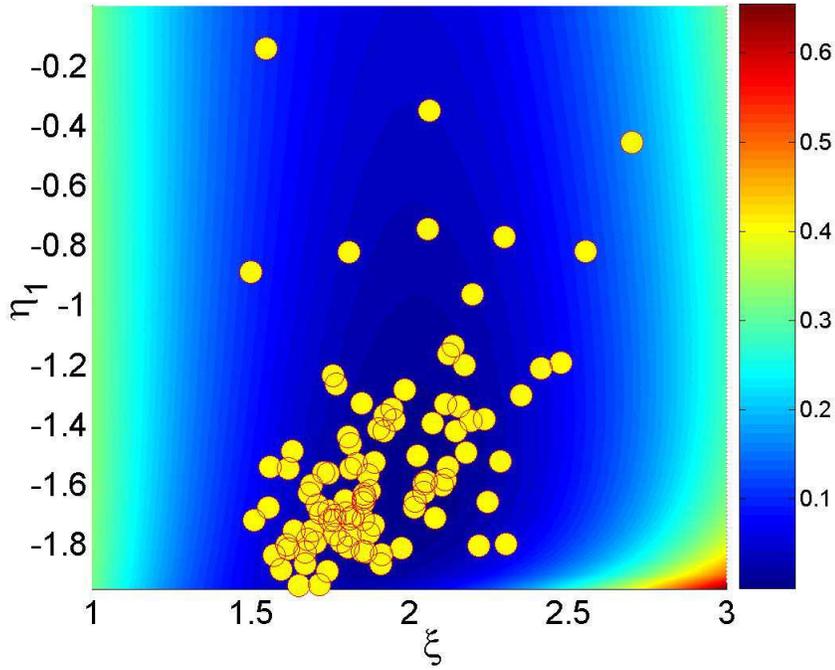}}
    \subfigure[MLE]{\label{fig:gauss_nll}\includegraphics[scale=0.3]{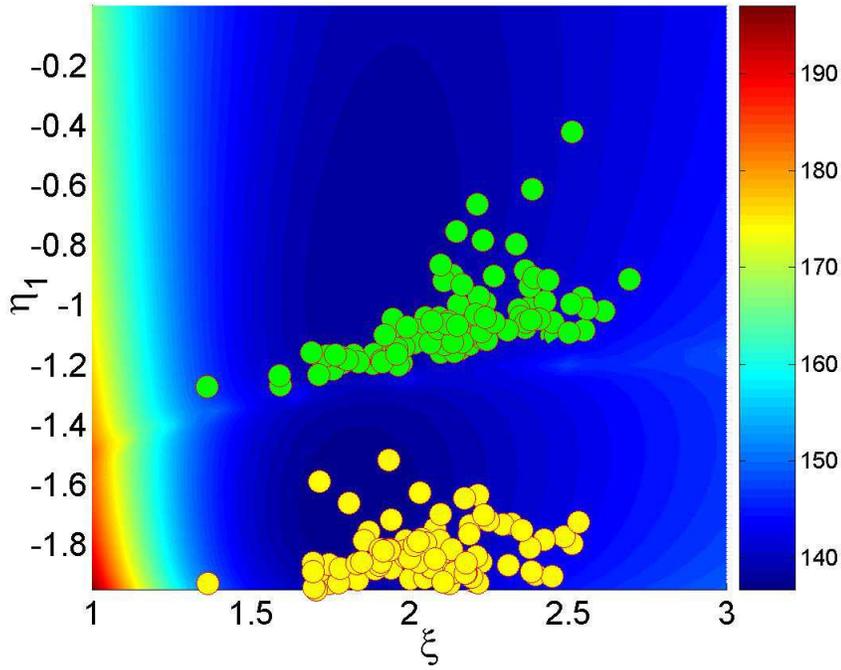}}
  \end{center}
  \caption{Top view of (a) the distance metric and (b) the negative log-likelihood
  function of one realization of the time series with the Gaussian covariance dependence
  and the parameters values $(\sigma,b)=(10,3)$ for $N=100$, projected onto the $(\eta_1,\xi)$ plane.
  The yellow circles represent the locations of optimal values of ($\eta_1^{*},\xi^{*}$),
  obtained from 100 different realizations, calculated by (a) MMoM and (b) MLE.
  The green circles in (b) represent solutions stuck in local minima.}
  \label{fig:gauss_FF}
\end{figure}

In Fig. \ref{fig:gauss_dm}, the DM function of one realization
($N=100$) is projected onto the parameter space $(\eta_1,\xi)$. The
optimal values $\eta_1^{*},\xi^{*}$, obtained from the $100$
realizations are also marked on this plane. The DM has a single
minimum in the parameter plane. The estimates of  $\eta_1^{*}$ tend
to spread away from the permissibility boundary at $\eta_1=-2,$
producing a skewed distribution with relatively large variance. A
similar plot of the NLL function, Fig. \ref{fig:gauss_nll}, exhibits
two local minima. The optimal values $\eta_1^{*},\xi^{*}$, shown in
green are obtained using the initial guess $\eta_1^{(0)}=-1$, while
those in yellow (corresponding to the global minimum) are obtained
with $\eta_1^{(0)}=-1.7$.

\begin{figure}[htbp]
  \begin{center}
\rotatebox{0}{ \resizebox{10cm}{!}{
\includegraphics{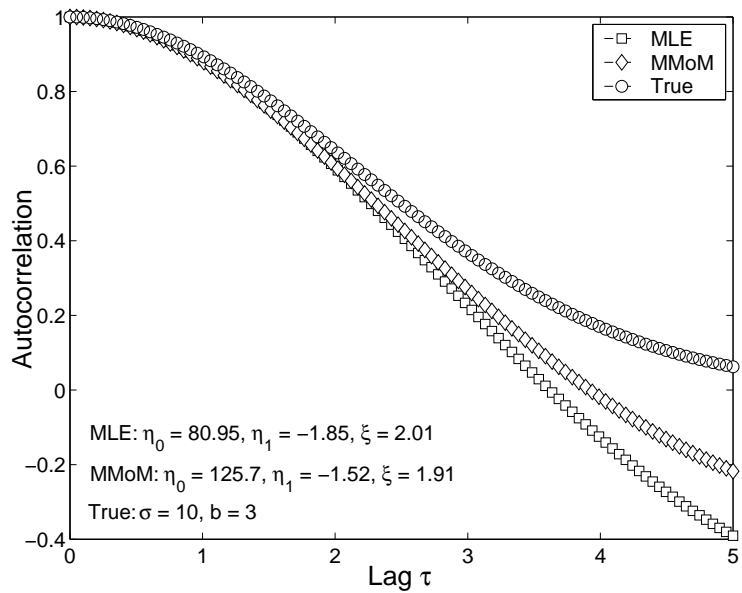}}}
  \end{center}
\caption{Plots of the true correlation function with the Gaussian
dependence and the true parameters $\sigma = 10$, $b = 3$ (circles),
and the Spartan ones using the parameter sets estimated by MMoM
(diamonds) and MLE (squares).} \label{fig:gauss_correl}
\end{figure}

For interpolation purposes, the differences in the parameter sets
obtained by means of the two methods are practically negligible. In
Fig. \ref{fig:gauss_correl}, the Spartan correlation functions
obtained with the two sets of parameters are compared with the true
(Gaussian) correlation. Near the origin and at intermediate
distances, which are crucial for the interpolation, the correlation
functions almost coincide. The negative values of $\eta_1^{*}$
causes a negative hole in the Spartan correlation at larger
distances. This hole is only observed in the Gaussian case due to
the short range of the constraints.

Considering the optimization statistics, the value of the DM
function ($\sim 10^{-21}$) means an excellent match between  the
sample and stochastic constraints. The MMoM needs more iterations
than the MLE to converge. However, the MMoM CPU time is much lower
and almost insensitive to the domain size, while the MLE CPU time
grows rapidly with $N$. Since the number of iterations is
insensitive to $N$ in both MLE and MMoM, the CPU time per iteration
shows similar behavior with increasing $N$ as the total CPU time.
For example, for $N=1 \,000$ the MMoM  is $3 \,965$ times faster
than the MLE. In the case of the Gaussian model reasonable CPU times
are attained for the domain sizes considered. However, the
optimization slows down considerably for the other covariance
models. This is due to the presence of a long flat valley in  the DM
and the NLL functions, which extends to large $\eta_1 \approx 10^5$.
Convergence requires a high number of iterations, $O(10^3)$, which,
for larger $N,$ is easily manageable by MMoM but too expensive
computationally in the case of MLE. Hence, in the following we will
use MMoM for parameter inference.

\subsection{Results: Interpolation}
\begin{figure}[htbp]
  \begin{center}
\rotatebox{0}{ \resizebox{10cm}{!}{
\includegraphics{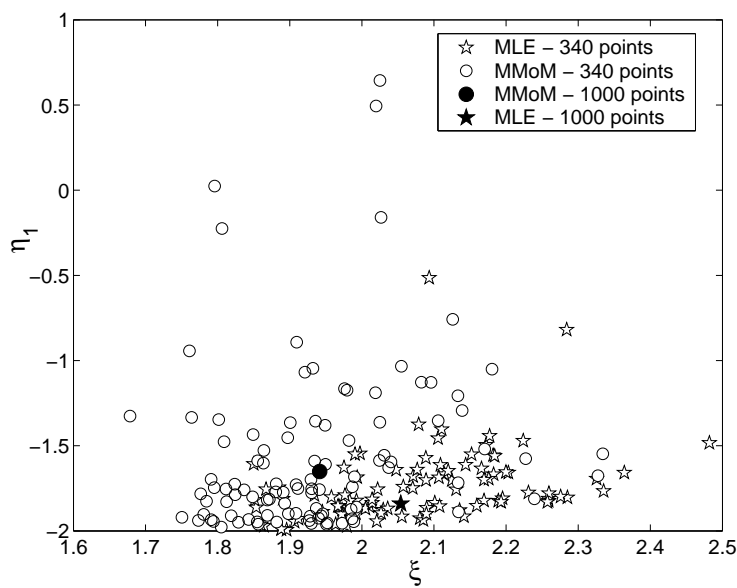}}}
  \end{center}
\caption{Optimal $\eta_1^{*},\xi^{*}$ estimates from  complete
series of 1\, 000 points calculated by MLE (filled star) and MMoM
(filled circle). Also, distribution of the optimal
$\eta_1^{*},\xi^{*}$ estimates based on the training set of 340
points calculated by MLE (empty stars) and MMoM (empty circles).}
\label{fig:gauss_xi-eta1}
\end{figure}

In the following we investigate the problem of interpolation
(missing data prediction). Let us denote by $p$ the degree of
thinning, i.e., the percentage of the missing data. We select one
time series with $N=1\, 000$ per covariance model and we generate
$100$ different partitions into a validation set, containing  $660$
points at random, and a training set, containing  $340$ points
(i.e., $p \approx 0.66.$) For comparison purposes, in Fig.
\ref{fig:gauss_xi-eta1}, we show the $\eta_1^{*},\xi^{*}$ estimates
for $p=0$ (full data) and the distribution of the
$\eta_1^{*},\xi^{*}$ values for $p=0.66$, calculated by both the MLE
and the MMoM.

We use both  KWP (with search neighborhood extending over the entire
series) and the Spartan predictor. The covariance function is
calculated using the asymptotic formulas given by Eqs.
(\ref{eq:spart-cov}). For the analysis of prediction performance,
the validation points are segregated into $9$ different categories,
based on the number of nearest and next-nearest neighbors that
belong in the training set. For example, the category $(i,j), \,
i,j=0,1,2,$ includes those validation points for which $i$ nearest
neighbors and $j$ second-nearest neighbors are in the training set.

The following statistics are calculated, for each category
separately, and over all the validation points: mean absolute error
(MAE), mean relative error (MRE), mean absolute relative error
(MARE), and root mean square error (RMSE). The calculations are performed
for the four covariance models and results for the Gaussian and
the exponential cases are shown in Tables \ref{tab:predict_gauss}
and \ref{tab:predict_expo}. Overall, validation
points with more data in their interaction neighborhood display
smaller errors than those with fewer data. In the Gaussian and
the Whittle-Mat\'{e}rn cases the KWP performs slightly better than the SP,
while in the exponential and the spherical cases there
is no significant difference between the two methods.

%
%
%
%
%
%
%

\begin{table}[h]
\caption{SP and KWP performance comparison on synthetic data with
Gaussian  covariance and parameters: mean $m = 50$, standard
deviation $\sigma = 10$ and correlation length $b = 3$. The Spartan
parameters were inferred from a training set of $340$ points. The
errors were calculated from the validation set of $660$ points.}
\label{tab:predict_gauss}
\begin{center}
\begin{tabular}{llrrrrrrrrrr}
\hline
&   &      (2,2) &      (1,2) &      (0,2) &      (0,1) &      (0,0) &      (1,1) &      (1,0) &      (2,0) &      (2,1) &      Total \\
\hline
MAE &        SP &      0.40 &      1.17 &      3.30 &      5.36 &      7.98 &      1.89 &      2.93 &      0.90 &      0.64 &      3.83 \\
&       KWP &      0.27 &      0.96 &      3.17 &      5.15 &      7.85 &      1.72 &      2.86 &      0.86 &      0.52 &      3.69 \\
\hline
MARE [\%] &        SP &      0.8 &      2.5 &      7.0 &      11.5 &      17.4 &      4.0 &      6.2 &      1.9 &      1.3 &      8.2 \\
&       KWP &      0.6 &      2.0 &      6.7 &      10.9 &      17.1 &      3.6 &      6.0 &      1.8 &      1.1 &      7.9 \\
\hline
MRE [\%] &        SP &     $-$0.1 &     $-$0.2 &     $-$1.1 &     $-$1.7 &     $-$4.0 &     $-$0.4 &     $-$0.5 &     $-$0.2 &     $-$0.1 &     $-$1.3 \\
&       KWP &      0.0 &     $-$0.1 &     $-$0.9 &     $-$1.5 &     $-$3.9 &     $-$0.3 &     $-$0.5 &     $-$0.1 &     $-$0.1 &     $-$1.3 \\
\hline
RMSE &        SP &      0.49 &      1.46 &      4.13 &      7.25 &     10.28 &      2.42 &      3.82 &      1.15 &      0.84 &      6.03 \\
&       KWP &      0.32 &      1.21 &      3.98 &      6.43 &      9.62 &      2.16 &      3.61 &      1.10 &      0.66 &      5.51 \\
\hline
\end{tabular}

\end{center}
\end{table}

\begin{table}[h]
\caption{SP and KWP performance comparison on synthetic data with
exponential  covariance and parameters: mean $m = 50$, standard
deviation $\sigma = 10$ and correlation length $b = 5$. The Spartan
parameters were inferred from a training set of $340$ points. The
errors were calculated from the validation set of $660$ points.}
\label{tab:predict_expo}
\begin{center}
\begin{tabular}{llrrrrrrrrrr}
\hline
&  &      (2,2) &      (1,2) &      (0,2) &      (0,1) &      (0,0) &      (1,1) &      (1,0) &      (2,0) &      (2,1) &      Total \\
\hline
MAE &        SP &      3.81 &      4.08 &      4.87 &      5.72 &      6.76 &      4.33 &      4.49 &      3.60 &      3.66 &      5.03 \\
\hline
&       KWP &      3.84 &      4.11 &      4.87 &      5.71 &      6.75 &      4.36 &      4.49 &      3.61 &      3.68 &      5.04 \\
\hline
MARE [\%] &        SP &      8.0 &      8.6 &      10.4 &      12.3 &      14.6 &      9.2 &      9.6 &      7.6 &      7.8 &      10.8 \\
\hline
&       KWP &      8.0 &      8.7 &      10.4 &      12.3 &      14.6 &      9.3 &      9.6 &      7.6 &      7.8 &      10.8 \\
\hline
MRE [\%] &        SP &     $-$0.3 &     $-$1.1 &     $-$1.7 &     $-$2.4 &     $-$3.0 &     $-$1.3 &     $-$1.6 &     $-$0.8 &     $-$1.0 &     $-$1.8 \\
\hline
&       KWP &     $-$0.2 &     $-$1.1 &     $-$1.7 &    $-$2.3 &     $-$3.0 &     $-$1.2 &     $-$1.5 &     $-$0.7 &     $-$1.0 &     $-$1.8 \\
\hline
RMSE &        SP &      4.58 &      5.11 &      6.08 &      7.18 &      8.46 &      5.38 &      5.59 &      4.46 &      4.55 &      6.44 \\
\hline
&       KWP &      4.62 &      5.15 &      6.08 &      7.17 &      8.46 &      5.40 &      5.59 &      4.46 &      4.58 &      6.44 \\
\hline
\end{tabular}

\end{center}
\end{table}

\section{Case Study II: Aerosol concentration data}
\label{sec:real} Next, we consider an application of the Spartan
predictor to a time series consisting of concentration measurements
of atmospheric aerosol PM2.5 particles (i.e., particles of
aerodynamic diameter smaller than $2.5 \mu m$). These measurements
were sampled on the grounds of the Technical University of Crete
using the aerosol monitor DustTrak, over a period of almost $10$
days in June 2006. The original time series consists of $2 \,831$
mean
 concentration values, measured at $5$ minute intervals.
The correlations in the observed series span over large distances
(measured in the lag of $\alpha = 5$ minutes). In order to make the
interpolation task more challenging, we reduced the length of the
series by non-overlapping clustering the original data using
averages over $\alpha = 40$ minute intervals. This resulted in a
coarse-grained time series of $353$ data points that will be used
below. The data fail the Kolmogorov-Smirnov normality test at the
$5\%$ level. Nonetheless, in the following we will model the series
with a second-order stationary FGC Spartan random process. The
motivation for this approach is that reasonable interpolation
performance can be obtained by matching the short-range
correlations.

\begin{figure}[htbp]
  \begin{center}
\rotatebox{0}{ \resizebox{10cm}{!}{
\includegraphics{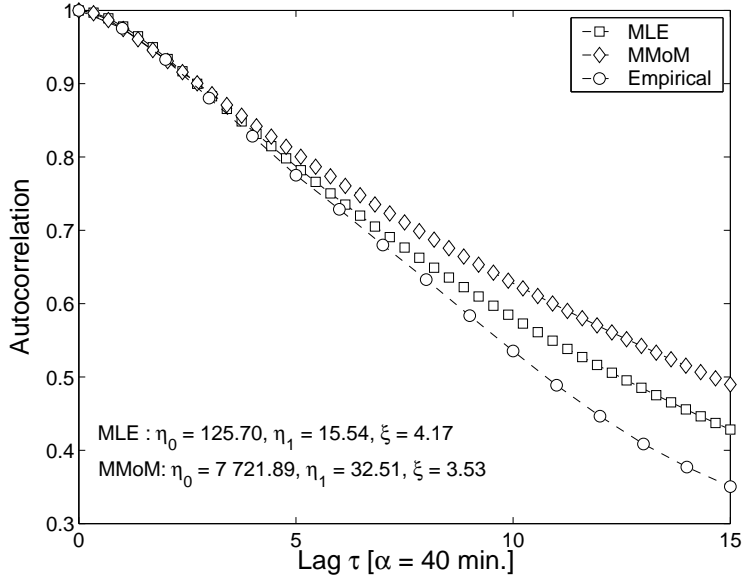}}}
  \end{center}
\caption{Plots of the empirical correlation function of the aerosol
concentration data (circles), and the Spartan ones using the
parameters estimated by MMoM (diamonds) and MLE (squares).
The lag is measured in units of the sampling step $\alpha = 40$
min.} \label{fig:real_correl}
\end{figure}

\begin{table}[h]
\caption{Spartan model parameters $\eta_0^{*}$, $\eta_1^{*}$ and
$\xi^{*}$,  number of optimization iterations, $N_{it}$, the
optimization CPU time, $T_{cpu}$, and the final value $F^{*}$ of the
objective function (NLL for MLE and DM for MMoM), calculated from
the complete aerosol concentration time series by MLE and MMoM.}
\label{tab:par_inf_real}
\begin{center}
\begin{tabular}{lrrrrrr}
\hline
           & $\eta_0^{*}$ & $\eta_1^{*}$ & $\xi^{*}$ & $N_{it}$ & $T_{cpu}[s]$ &  $F^{*}$ \\
\hline
       MLE &    5\ 213.90 &      15.54 &       4.17 &      76.00 &      16.03 &     585.89 \\
      MMoM &    7\ 721.90 &      32.51 &       3.53 &     182.00 &       0.39 &     2E$-29$ \\
\hline
\end{tabular}

\end{center}
\end{table}

The Spartan parameters based on the complete time series are shown
in Table~\ref{tab:par_inf_real}). The number of iterations involved
in the MMoM is more than twice as high as that for the MLE, but the
CPU time is $41$ times faster. The final DM value of $F^{*}=2\,
10^{-29}$ indicates excellent matching between the sample and
stochastic constraints. In Fig.~\ref{fig:real_correl}, we compare
the correlation functions based on the estimated Spartan parameters
with the experimental correlation. Very good agreement of the
experimental and estimated functions at short time ranges is
observed.

\begin{table}[h]
\caption{SP and KWP performance comparison on the aerosol
concentration data.  The Spartan parameters were inferred by MLE,
using the training set of 121 points and the errors were calculated
on the validation set of the remaining 232 points. Category $(i,j)$
includes points with $i$ nearest and $j$ next-nearest neighbors from
the training set.} \label{tab:predict_real_mle}
\begin{center}
\begin{tabular}{llrrrrrrrrr}
\hline
&  &      (2,2) &      (1,2) &      (0,2) &      (0,1) &      (0,0) &      (1,1) &      (1,0) &      (2,0) &      (2,1) \\
\hline
MAE &        SP &      1.75 &      2.67 &      3.85 &      5.30 &      8.15 &      2.88 &      3.24 &      2.05 &      2.03 \\
&       KWP &      1.80 &      2.68 &      3.87 &      5.32 &      8.10 &      2.90 &      3.25 &      2.06 &      2.06 \\
\hline
MARE [\%] &        SP &      4.3 &      6.6 &      9.3 &      12.5 &     19.1 &      6.8 &      7.5 &      4.5 &      4.9 \\
&       KWP &      4.4 &      6.7 &      9.4 &      12.6 &      19.0 &      6.9 &     7.6 &      4.5 &      5.0 \\
\hline
MRE [\%]  &      SP &     $-$0.4 &     $-$1.1 &     $-$1.4 &     $-$2.6 &     $-$5.2 &     $-$0.9 &     $-$0.9 &     $-$0.2 &     $-$0.8 \\
&       KWP &     $-$0.4 &     $-$0.9 &     $-$1.3 &     $-$2.5 &     $-$5.0 &     $-$0.8 &     $-$0.8 &     $-$0.2 &     $-$0.6 \\
\hline
RMSE     &   SP &      2.15 &      3.72 &      5.14 &      7.44 &     11.04 &      4.23 &      4.65 &      2.92 &      2.93 \\
&       KWP &      2.19 &      3.73 &      5.12 &      7.46 &     10.97 &      4.25 &      4.65 &      2.92 &      2.98 \\
\hline
\end{tabular}
\end{center}
\end{table}

\begin{figure}[htp]
  \begin{center}
    \subfigure[MMoM]{\label{fig:real_dm}\includegraphics[scale=0.3]{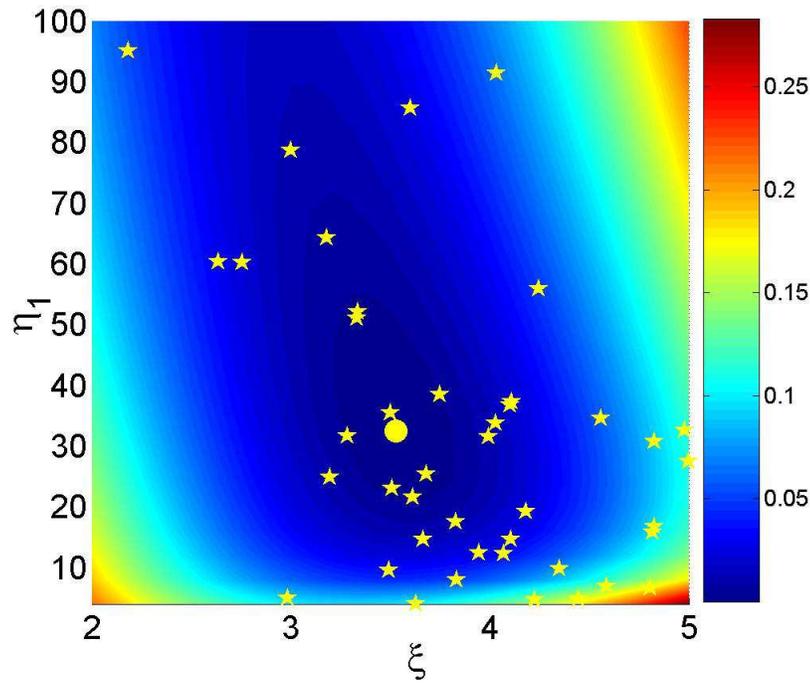}}
    \subfigure[MLE]{\label{fig:real_nll}\includegraphics[scale=0.3]{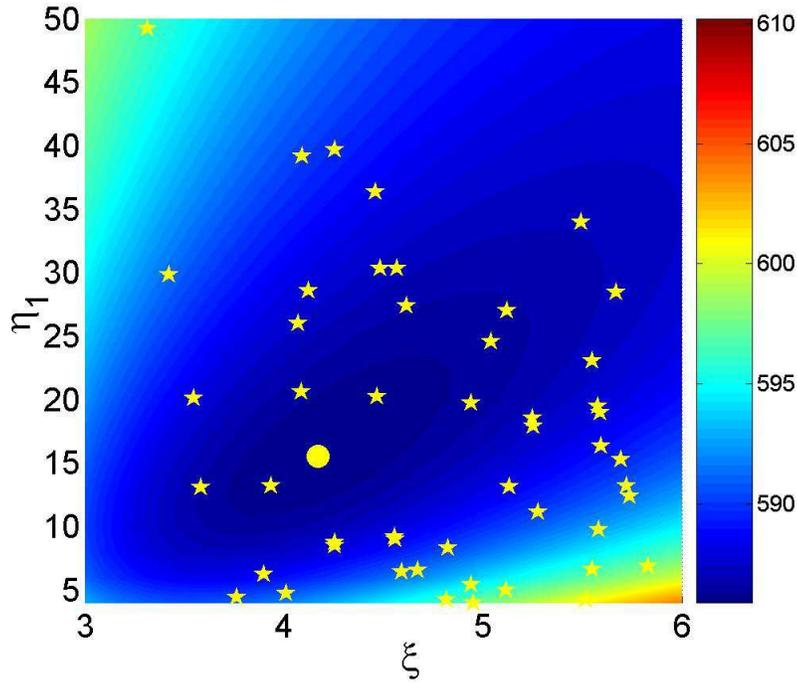}}
  \end{center}
  \caption{Top view of (a) the distance metric and (b) the negative log-likelihood function of the complete aerosol concentration time series, projected onto the $(\eta_1,\xi)$ plane. The yellow circles represent the locations of optimal values of ($\eta_1^{*},\xi^{*}$), obtained from the complete data and the yellow stars those obtained from 100 different realizations of the training set of 121 points, calculated by (a) MMoM and (b) MLE (a few points are out of the scale range).}
  \label{fig:real_ff}
\end{figure}

For the interpolation, the data are partitioned into a training set
that  involves $121$ randomly selected times, and a validation set
including the remaining $232$ values. The parameter inference on the
training set is performed by both MMoM and MLE. There is no
multimodality in either NLL or DM surface. However, as shown in
Fig.~\ref{fig:real_ff}, there is considerable scatter in the
parameter estimates obtained from different training set
realizations.

We calculate the same interpolation error measures as for the
synthetic data. The results are shown in Table
\ref{tab:predict_real_mle}, using MLE estimates of the Spartan
parameters. The differences between the SP and the KWP are minimal.
The estimates obtained using the MLE based Spartan parameters are
slightly better than the ones based on the MMoM. For comparison, the
total mean absolute relative error of the MLE-based estimates is
only $0.8 \%$ (using the Spartan predictor) and $0.9 \%$ (using the
KWP) smaller than the MMoM one.

\begin{figure}[htbp]
  \begin{center}
\rotatebox{0}{ \resizebox{10cm}{!}{
\includegraphics{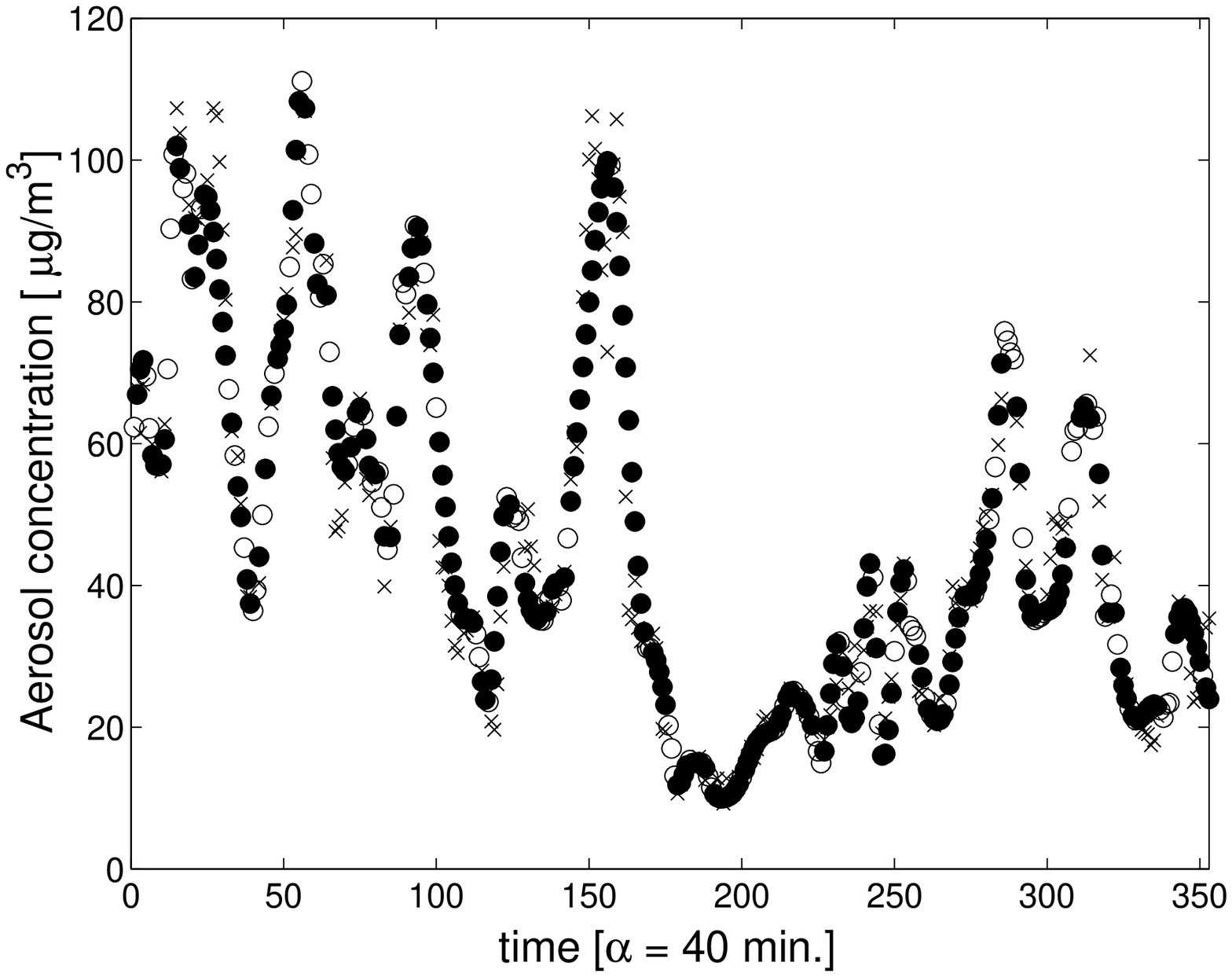}}}
  \end{center}
\caption{Aerosol concentration time series reconstruction for one
training set realization of 121 points. The data in training set
are marked by empty circles. The data in the validation set are
marked by crosses. The  estimates at the validation set
locations are marked by filled circles. The time unit is $\alpha=40$
minutes.} \label{fig:real_reconstr}
\end{figure}

In Fig. \ref{fig:real_reconstr}, we illustrate the missing data
reconstruction based on the Spartan predictor. As the data in Table
\ref{tab:predict_real_mle} already suggest, the largest deviations of
the estimates from the actual
data are seen in the cases of "isolated" points with no training
data in their interaction neighborhood, i.e., in the category (0,0).
The contribution to the total errors from such points can be
considerable if the missing data represent a large portion of the
entire series. In Table \ref{tab:predict_real_t}, the total errors
are shown versus the percentage of the missing data. The mean values
of the correlation coefficient ($R$) between the estimated and actual
values at the missing points is also included. High correlation
values $(\approx 95\%)$ are observed even for $p=0.66$.

\begin{table}[h]
\caption{SP and KWP performance comparison based on the aerosol
concentration  time series for different sizes of missing data. The
Spartan parameters are inferred by MMoM using training sets of
$(1-p)*N$ points. The errors are calculated based on the remaining
$p*N$ points, for $N=353$ and $p=0.66,0.6,0.4, 0.2$.}
\label{tab:predict_real_t}
\begin{center}
\begin{tabular}{llrrrr}
\hline
&   $p$ &       0.66 &       0.60 &       0.40 &       0.20 \\
\hline
MAE &      SP &      4.66 &      3.86 &      2.81 &      2.05 \\
&      KWP &      4.69 &      3.85 &      2.81 &      2.00 \\
\hline
MARE [\%] &      SP &      11.1 &      9.1 &      6.6 &      5.6 \\
&      KWP &      11.2 &      9.1 &      6.6 &      5.4 \\
\hline
MRE [\%] &      SP &     $-$2.6 &     $-$1.7 &     $-$0.7 &     $-$1.2 \\
&      KWP &     $-$2.5 &     $-$1.6 &     $-$0.8 &     $-$1.0 \\
\hline
RMSE &      SP &      7.18 &      6.00 &      4.34 &      3.07 \\
&      KWP &      7.25 &      5.98 &      4.34 &      2.97 \\
\hline
$R$ &      SP &      0.95 &      0.97 &      0.99 &      0.99 \\
&      KWP &      0.95 &      0.97 &      0.99 &      0.99 \\
\hline
\end{tabular}

\end{center}
\end{table}

\section{Conclusions}
\label{concl} We present a framework for the analysis of Gaussian,
stationary time series based on Spartan random processes.  In this
framework, the temporal dependence is determined from
`pseudo-energy' functionals.  The modified method of moments (MMoM)
is proposed for Spartan parameter estimation. Its main advantages
are low computational complexity (high speed) even for large sample
sizes. Temporal interpolation is formulated by maximizing the
conditional probability density function of the Spartan process,
based on the available data. The method is tested with synthetic
data and with a time series of atmospheric PM2.5 aerosol
concentration. The Spartan interpolator is shown to perform
similarly to the standard Kolmogorov-Wiener predictor at reduced
computational cost.

In time series modeling, it is often necessary to treat  the model
parameters as time-dependent and to estimate them continuously in
the "moving window" fashion. This approach  accounts for lack of
stationarity, which is a common feature in meteorological data. The
"moving window" approach has been recently applied to the automatic
mapping of rainfall data, using maximum likelihood for parameter
inference and universal kriging as the interpolator~\citep{pard05}.
Windows are typically overlapping and of varying size. In each
window, parameter inference and model selection (based on the Akaike
Information Criterion) are conducted, followed by interpolation.
Spartan random processes have definite advantages for application in
the moving window framework. First, model selection can be bypassed
thanks to the flexibility of the SRP (compared with two or three
parameter covariance models). Second, the computational efficiency
of Spartan parameter inference and interpolation would allow for
iterative methods of window size adjustment that will properly
account for the physical conditions at the local scale.

\section*{Acknowledgements}
This research project has been supported by a Marie Curie Transfer
of Knowledge Fellowship of the European Community's Sixth Framework
Programme under contract number MTKD-CT-2004-014135.

The atmospheric aerosol data have been kindly provided by Dr. J.
Ondr\'{a}\v{c}ek and Dr. M. Lazaridis (Department of Environmental
Engineering, Technical University of Crete, Chania, GR 73100).

\newpage

{\bf List of figures}

{\bf Fig 1:} Some trivial time series, consisting of only 3 points,  for
which equivalence between SP and KWP can be shown analytically. The
circles denote the known data and the crosses the prediction
points.

{\bf Fig 2:} Top view of (a) the distance metric and (b) the negative log-likelihood
function of one realization of the time series with the Gaussian covariance dependence
and the parameters values $(\sigma,b)=(10,3)$ for $N=100$, projected onto the $(\eta_1,\xi)$ plane.
The yellow circles represent the locations of optimal values of ($\eta_1^{*},\xi^{*}$),
obtained from 100 different realizations, calculated by (a) MMoM and (b) MLE.
The green circles in (b) represent solutions stuck in local minima.
  
{\bf Fig 3:} Plots of the true correlation function with the Gaussian
dependence and the true parameters $\sigma = 10$, $b = 3$ (circles),
and the Spartan ones using the parameter sets estimated by MMoM
(diamonds) and MLE (squares).  

{\bf Fig 4:} Optimal $\eta_1^{*},\xi^{*}$ estimates from  complete
series of 1\, 000 points calculated by MLE (filled star) and MMoM
(filled circle). Also, distribution of the optimal
$\eta_1^{*},\xi^{*}$ estimates based on the training set of 340
points calculated by MLE (empty stars) and MMoM (empty circles).

{\bf Fig 5:} Plots of the empirical correlation function of the aerosol
concentration data (circles), and the Spartan ones using the
parameters estimated by MMoM (diamonds) and MLE (squares).
The lag is measured in units of the sampling step $\alpha = 40$
min.

{\bf Fig 6:} Top view of (a) the distance metric and (b) the negative log-likelihood function of the complete aerosol concentration time series, projected onto the $(\eta_1,\xi)$ plane. The yellow circles represent the locations of optimal values of ($\eta_1^{*},\xi^{*}$), obtained from the complete data and the yellow stars those obtained from 100 different realizations of the training set of 121 points, calculated by (a) MMoM and (b) MLE (a few points are out of the scale range).

{\bf Fig 7:} Aerosol concentration time series reconstruction for one
training set realization of 121 points. The data in training set
are marked by empty circles. The data in the validation set are
marked by crosses. The  estimates at the validation set
locations are marked by filled circles. The time unit is $\alpha=40$
minutes.

\end{document}